\documentclass[11pt]{article}
\usepackage{amssymb}
\usepackage{citesort}
\usepackage{url}
\usepackage{a4}

\newcommand{\tg}{\tilde g}
\newcommand{\riemgz}{g_0}

\renewcommand{\hbar}{{\overline \riemgz}}

\newcommand{\nablash}{\nabla{\kern -.75 em
     \raise 1.5 true pt\hbox{{\bf/}}}\kern +.1 em}
\newcommand{\Deltash}{\Delta{\kern -.69 em
     \raise .2 true pt\hbox{{\bf/}}}\kern +.1 em}
\newcommand{\Rslash}{R{\kern -.60 em
     \raise 1.5 true pt\hbox{{\bf/}}}\kern +.1 em}

\newcommand{\mcM}{{\mycal M}}
\newcommand{\bmcM}{\,\,\,\,{\widetilde{\!\!\!\!\mycal M}}}
\newcommand{\tM}{\bmcM}

\newcommand{\bea}{\begin{eqnarray}}
\newcommand{\beaa}{\begin{eqnarray*}}
\newcommand{\bean}{\begin{eqnarray}\nonumber}

\newcommand{\bel}[1]{\begin{equation}\label{#1}}
\newcommand{\beal}[1]{\begin{eqnarray}\label{#1}}
\newcommand{\beadl}[1]{\begin{deqarr}\label{#1}}
\newcommand{\eeadl}[1]{\arrlabel{#1}\end{deqarr}}
\newcommand{\eeal}[1]{\label{#1}\end{eqnarray}}
\newcommand{\eead}[1]{\end{deqarr}}
\newcommand{\eea}{\end{eqnarray}}
\newcommand{\eeaa}{\end{eqnarray*}}

\newcommand{\be}{\begin{equation}}
\newcommand{\ee}{\end{equation}}

\newcommand{\eq}[1]{\eqref{#1}}

\DeclareFontFamily{OT1}{rsfs}{}
\DeclareFontShape{OT1}{rsfs}{m}{n}{ <-7> rsfs5 <7-10> rsfs7 <10->
rsfs10}{} \DeclareMathAlphabet{\mycal}{OT1}{rsfs}{m}{n}
\def\scri{{\mycal I}}%
\def\Scri{\scri}

\usepackage{amsmath} 

\let\ssection=\section

\renewcommand{\section}{\setcounter{equation}{0}\ssection}

\def \Reel{\mathbb{R}}

\def \R {\Reel}

\newcounter{mnotecount}[section]

\renewcommand{\themnotecount}{\thesection.\arabic{mnotecount}}

\newcommand{\mnote}[1]
{\protect{\stepcounter{mnotecount}}$^{\mbox{\footnotesize $
\bullet$\themnotecount}}$ \marginpar{
\raggedright\tiny\em $\!\!\!\!\!\!\,\bullet$\themnotecount: #1} }

\newcommand{\rmnote}[1]{}

\begin{document}
\title{On solutions of the vacuum Einstein equation in the radiation regime}
\author{
Piotr T. Chru\'sciel\thanks{Partially supported by a Polish
Research Committee grant; email \protect\url{
piotr@gargan.math.univ-tours.fr}} \\
 D\'epartement de
math\'ematiques\\ Facult\'e des Sciences\\ Parc de Grandmont\\
F37200 Tours, France}
\date{}

\maketitle


\begin{abstract} We review recent results by the author, in collaboration with Erwann
Delay, Olivier Lengard, and Rafe Mazzeo, on existence and
properties of space-times with controlled asymptotic behavior at
null infinity.
\end{abstract}

The standard description of gravitational radiation proceeds as
follows: one considers space-times $(\mcM,g)$ which can be
conformally completed by adding a boundary $\partial \mcM=\scri$,
so that an appropriate conformal rescaling of $g$ leads to a
metric which extends by continuity to a Lorentzian metric $\tg$ on
the new manifold with boundary $\tM:=\mcM\cup\scri$. This raises
several questions:  do there exist non-trivial vacuum space-times
in which this can be done? does this prescription cover all
radiating space-times? or at least all interesting ones? is this
the right way to proceed anyway? In this talk I will review some
recent progress concerning those questions.

Recall~\cite{penrose:scri} that a space-time is called
\emph{asymptotically simple} if the above conformal completion
$(\tM,\tg)$ is smooth, and if every null geodesic of $(\mcM,g)$
has precisely two end points on $\scri$. It has been an open
question whether there exist any vacuum asymptotically space-times
other than the Minkowski one. A celebrated theorem of
Christodoulou and Klainerman~\cite{Christodoulou:Klainerman}
proves existence of a large family of space-times which are close
to being asymptotically simple: The Christodoulou-Klainerman
metrics are geodesically complete, and admit conformal
completions. However, the differentiability properties of the
conformally rescaled metrics are poorly controlled. This last
issue does play a role in the theory, as the properties of
$\scri$'s with low differentiability are rather different from
those of the smooth ones. For instance, the peeling properties of
the gravitational field, which are sometimes considered as a
characteristic feature of gravitational radiation, are different
for conformal completions which are, or which  are not, of $C^3$
differentiability class  ({\em cf., e.g.}\/~\cite{AC,TodCargese}).
Further, while it is rather likely that null geodesics will also
have precisely two end points on $\scri$ for the
Christodoulou-Klainerman space-times, no analysis of this question
is known to the author. In conclusion, it is not {\em a priori}
clear whether any of the spaces-times constructed by
Christodoulou-Klainerman are asymptotically simple in the original
sense of Penrose.\footnote{The space-times of~\cite{ChDelay2}
described below do actually belong to the Christodoulou-Klainerman
class, so it is known in retrospect that some of the
Christodoulou-Klainerman space-times will fit the Penrose scheme.}

The main difficulty here can be traced back to the following
question: how to control the asymptotic properties of the
gravitational field near $\scri$ in terms of the asymptotic
properties of the initial data in a neighborhood of $i^0$. We note
a recent important paper of Friedrich~\cite{Friedrich:2002ru}
which provides the first result in this direction for the
linearised spin-2 equations. However, the transition from the
linearised theory to the non-linear one does not appear to be
straightforward, so that work remains to be done in order to go
from~\cite{Friedrich:2002ru} to a full answer to the question.

A pioneering construction of non-trivial, physically interesting,
asymptotically simple space-times is due to Cutler and
Wald~\cite{CutlerWald}, for gravitation interacting with an
electromagnetic field. Cutler and Wald's idea is to construct a
non-trivial family of initial data on $\R^3$ for the coupled
Einstein-Maxwell equations which are exactly Schwarzschildian
outside of a ball, and which are as close to Minkowski initial
data as desired. The main interest of such data stems from the
fact that the evolution of the initial data set provides a
space-time which is exactly the Schwarzschild one near $i^0$.  For
data small enough one then obtains an asymptotically simple
space-time by invoking a stability theorem of
Friedrich~\cite{HelmutJDG}.

To repeat this argument in vacuum all one needs  is appropriate
initial data. In a recent work with Erwann Delay~\cite{ChDelay2}
we have been able to construct such data, using a variation of an
important construction of Corvino and
Schoen~\cite{CorvinoSchoen,Corvino}. In~\cite{ChDelay2} we
consider vacuum, time symmetric ($K_{ij}\equiv 0$) initial data on
a ball $B(0,1)\subset \R^3$ satisfying the parity condition
\bel{pc} g_{ij}(\vec x) = g_{ij}(-\vec x)\;.\ee (Large families of
such metrics can be constructed using the conformal method.) We
show that all such metrics on $B(0,1)$ which are close enough to
Minkowski space-time can be extended to scalar flat metrics on
$\R^3$ which are exactly Schwarzschildian on $\R^3\setminus
B(0,2)$. As already mentioned, the technique is a gluing
construction along the lines of that of~\cite{Corvino}. The point
of the parity condition is the following: one of the steps of the
proof of the extension theorem requires an adjustment of the
centre of mass $\vec c$ of a tentative  Schwarzschildian
extension. This adjustment requires a translation of the
Schwarzschild metric by a vector $\vec c/m$. When $m$ gets small
the vector $\vec c/m$ could become very large, in which case the
construction breaks down. The parity condition \eq{pc} guarantees
that the centre of mass $\vec c$ is zero regardless of $m$, and
the problem with the division by the small parameter $m$ goes
away.

 Summarising: one starts with any one-parameter family of time
symmetric vacuum initial data on $B(0,1)$ satisfying \eq{pc}, such
that the initial data approach the trivial ones as the parameter
goes to zero. For values of the parameter small enough one can
extend the initial data to Schwarzschildian ones using the
extension technique above. Making the parameter smaller if
necessary one obtains an asymptotically
simple\footnote{In~\cite{ChDelay2} we asserted that the
construction leads to $C^k$ conformal completions, where $k$ can
be chosen at will, but finite. However, one can
show~\cite{ChDelay,Corvino} that if the starting metric on
$B(0,1)$ is smooth up-to-boundary, then things can be arranged so
that the resulting conformal completion will also be smooth.}
space-time using Friedrich's conformal system of
equations~\cite{Friedrich}.

Exactly the same construction allows one to make an extension in
which the Schwarzschild metric is replaced by any asymptotically
flat metric with non-zero mass $m$, satisfying the parity
condition \eq{pc}, in the following sense: let $g(1)$ be any
metric defined for $|\vec x|$ large enough, and consider the
family of metrics $g({\lambda})$ obtained by scaling,
$g(1)_{ij}(\vec x)\to g(\lambda)_{ij}(\vec x):=g(1)_{ij}(\lambda
\vec x)$. Then the mass $m(\lambda)$ of $g(\lambda)$ equals
$m/\lambda$. If the initial vacuum metric on $B(0,1)$ is close
enough to the flat one, then there exists a scalar flat extension
which coincides with $g(\lambda)$ on $\R^3\setminus B(0,2)$, for
some $\lambda$ large enough. This construction becomes most useful
when $g=g(1)$ satisfies the static constraint equations;
equivalently, when $g(1)$ arises from a static solution of the
vacuum Einstein equations. Since stationary vacuum metrics have a
smooth $\scri$~\cite{Dain:2001kn,Damour:schmidt}, one thus obtains
a reasonably large new family of asymptotically simple
space-times.

In~\cite{ChDelay2} yet another interesting application of the
Corvino-Schoen technique has been pointed out: the extension
theorem can be used to construct initial data sets for a
``many-black-hole" space-time, as follows: one chooses any number
of non-intersecting balls $B_i$, $i=1,\ldots,I$, which are
symmetrically distributed around the origin. To each of those
balls one assigns a small positive mass parameter $m_i$, again
symmetrically with respect to the origin. If the parameters $m_i$
are small enough, then~\cite{ChDelay2} one can find a vacuum
initial data set 
such that the initial Riemannian metric is exactly the (space)
Schwarzschild metric, centred\footnote{Here we have in mind the
conformally flat representation of the (space) Schwarzschild
metric, in which the "other infinity" of the Einstein-Rosen bridge
corresponds to the centre of the ball.} on the centre of $B_i$,
with mass $m_i$, within each of the balls $B_i$, and also exactly
the (space) Schwarzschild metric outside of a sufficiently large
ball, with some mass $m$ which is close to the sum of the masses
$m_i$. The fact that the metric is exactly Schwarzschildian in
each of the balls $B_i$ guarantees that there will be $I$
marginally trapped surfaces in the initial data set. Further, the
evolution of the metric in the domain of dependence of the $B_i$'s
will produce exactly a Schwarzschild metric there, with an
associated black hole region. The fact that the metric is exactly
Schwarzschildian outside of a large ball guarantees that one has
reasonably good control of the properties of the resulting
$\Scri$. Those facts put together allow one to
establish~\cite{ChMazzeo} (see also~\cite{Chmbh}) the following
properties of the maximal globally hyperbolic development
$(\mcM,g)$ of so-constructed initial data:
\begin{enumerate}
\item If the mass parameters are small enough, then the only
marginally trapped surfaces in the initial data set are the usual
minimal surfaces occurring in each of the Schwarzschildian balls.
Recall that a bounding marginally trapped surface within an
initial data set always encloses a black hole region, and a
non-connected {\em outermost}\/ marginally trapped surface is
usually interpreted as reflecting existence of a non-connected
black hole. \item Making the mass parameters smaller if necessary,
any configuration with two $B_i$'s will lead to a space-time in
which the intersection of the event horizon with the initial data
hypersurface will {\em not} be connected. Thus the initial data
surface does indeed contain two distinct black hole regions.
\end{enumerate}
We note that the paper~\cite{ChMazzeo} is the first one which
proves rigorously that some families of  vacuum initial data
contain  non-connected black hole regions. The proof again uses
Friedrich's stability results. Other existing
results~\cite{SchoenYaubh} or methods~\cite{KlainermanNicolo} do
not guarantee non-connectedness of the black hole.

The reader is referred to~\cite{ChDelay,CorvinoOberwolfach} for
further variations on the Corvino-Schoen technique.

 The space-times discussed so far
have smooth conformal completions, but they are  also extremely
special because of \eq{pc}. In fact, there have been various
indications
 that generic $\scri$'s will  not even be
$C^3$. First, this seems to occur in the Christodoulou-Klainerman
space-times (compare~\cite{Christodoulou:Rome}). Next, it has been
observed in~\cite{AC} that generic solutions of the constraint
equations on hyperboloids, constructed by the conformal method,
will not be $C^3$ at the surface where the hyperboloid intersects
$\scri$, even if the seed fields which enter the construction are
smooth there. It has been similarly observed in~\cite{ChMS}
(compare~\cite{GoldbergSoteriou,winicour:logarithmic}) that
generic initial data on outgoing null cones will not lead to
smooth $\scri$'s, even though the data induced on the initial
light cone are smooth at $\scri$. Now, those last two results
 leave open the
possibility that the resulting space-times will not have a smooth
$\scri$ for the simple reason that they will have no $\scri$ at
all. In recent work with O.~Lengard we have shown that this is not
the case, for large classes of hyperboloidal initial data.
In~\cite{ChLengardprep} one considers initial data in weighted
Sobolev spaces, with regularity at the boundary compatible with
that which is obtained in the conformal construction
of~\cite{AndChDiss,AC}. Roughly speaking, the weighting
corresponds to the following behavior of the derivatives of the
metric in terms of pseudo-Minkowskian coordinates:
\bel{ab}|\partial_u^i
\partial_r^j \partial_\theta^k \partial_\varphi^\ell
g_{\mu\nu}(u,r,\theta,\varphi)| \le C(i,j,k,\ell) r^{-3/2}\ee for
$j\ge 1$, and for $u=t-r$ near $u_0$.  This should be contrasted
with smooth conformal completions, which  would lead to a power
$r^{-1-j}$ in \eq{ab}. In those function spaces the derivatives of
the fields are too singular for the usual Friedrich evolution
theorems~\cite{Friedrich} to apply. It is shown
in~\cite{ChLengardprep} that the maximal globally hyperbolic
development of large classes of such initial data will possess a
conformal completion in a neighborhood of the initial data
hypersurface. The weighted Sobolev regularity (hence, also a
weighted $C^k$ regularity with a differentiability index $k$ lower
by two) of the metric is preserved during evolution.

Similar, and actually sharper, results can be proved for large
classes of semi-linear wave equations, and for the wave-map
equation, on Minkowski
space-time~\cite{ChLengardnwe,ChLengardBatz}. In particular for
such equations one can show that polyhomogeneity of the initial
data is preserved by evolution. We expect this to be also true for
Einstein equations, and we are hoping to settle this issue in the
near future.

\noindent {\sc Acknowledgements:} The author wishes to thank prof.
Ferrarese for the invitation to participate in this nice meeting,
in the beautiful surroundings of Elba.

\def\cprime{$'$}
\providecommand{\bysame}{\leavevmode\hbox
to3em{\hrulefill}\thinspace}
\providecommand{\MR}{\relax\ifhmode\unskip\space\fi MR }
\providecommand{\MRhref}[2]{%
  \href{http://www.ams.org/mathscinet-getitem?mr=#1}{#2}
} \providecommand{\href}[2]{#2}


\begin{thebibliography}{10}

\bibitem{AC}
L.~Andersson and P.T. Chru\'sciel, \emph{On ``hyperboloidal"
{C}auchy data for
  vacuum {E}instein equations and obstructions to smoothness of {S}cri},
  Commun.\ Math.\ Phys. \textbf{161} (1994), 533--568.

\bibitem{AndChDiss}
\bysame, \emph{On asymptotic behaviour of solutions of the
constraint equations
  in general relativity with ``hyperboloidal boundary conditions''}, Dissert.
  Math. \textbf{355} (1996), 1--100.

\bibitem{Christodoulou:Rome}
D.~Christodoulou, \emph{The global initial value problem in
general
  relativity}, Proceedings of the Ninth Marcel Grossman Meeting,
  \url{http://141.108.24.15:8000/}.

\bibitem{Christodoulou:Klainerman}
D.~Christodoulou and S.~Klainermann, \emph{On the global nonlinear
stability of
  {M}inkowski space}, Princeton University Press, Princeton, 1995.

\bibitem{Chmbh}
P.T. Chru\'{s}ciel, talk given at the Carg\`ese Summer School,
August 2002,
  online at \url{fanfreluche.math.univ-tours.fr}.

\bibitem{ChDelay2}
P.T. Chru\'{s}ciel and E.~Delay, \emph{Existence of non-trivial
asymptotically
  simple vacuum space-times}, Class. Quantum Grav. \textbf{19} (2002),
  L71--L79, gr-qc/0203053, erratum Class. Quantum Grav. {\bf 19} (2002), 3389.

\bibitem{ChDelay}
\bysame, \emph{On the general relativistic constraints operator in
weighted
  {S}obolev spaces}, in preparation, 2002.

\bibitem{ChLengardBatz}
P.T. Chru\'{s}ciel and O.~Lengard, \emph{Polyhomogeneous solutions
of wave
  equations in the radiation regime}, Journ\'ees Equations aux d\'eriv\'ees
  partielles, Nantes, 5-9 june, 2000 (X.~Saint Raymond~Eds. N.~Depauw,
  D.~Robert, ed.), \url{http://www.math.sciences.univ-nantes.fr/edpa/2000/},
  pp.~III--1 --- III--17.

\bibitem{ChLengardprep}
\bysame, \emph{Solutions of {Einstein equations polyhomogeneous at
Scri}}, in
  preparation.

\bibitem{ChLengardnwe}
\bysame, \emph{Solutions of wave equations in the radiating
regime},
  \url{http://www.phys.univ-tours.fr/~piotr/papers/batz/ls.html}.

\bibitem{ChMS}
P.T. Chru{\'s}ciel, M.A.H. MacCallum, and D.~Singleton,
\emph{Gravitational
  waves in general relativity. {XIV}: {B}ondi expansions and the
  ``polyhomogeneity'' of {S}cri}, Phil. Trans. Roy. Soc. London A \textbf{350}
  (1995), 113--141.

\bibitem{ChMazzeo}
P.T. Chru\'{s}ciel and R.~Mazzeo, \emph{On ``many black hole"
vacuum
  space-times},  (2002), \url{www.phys.univ-tours.fr/~piotr/papers/cs/ls.html}.

\bibitem{Corvino}
J.~Corvino, \emph{Scalar curvature deformation and a gluing
construction for
  the {E}instein constraint equations}, Commun.\ Math.\ Phys. \textbf{214}
  (2000), 137--189.

\bibitem{CorvinoOberwolfach}
\bysame, lecture in Oberwolfach, July 2000.

\bibitem{CorvinoSchoen}
J.~Corvino and R.~Schoen, \emph{Vacuum spacetimes which are
identically
  {S}chwarzschild near spatial infinity}, talk given at the Santa Barbara
  Conference on Strong Gravitational Fields, June 22-26, 1999,
  \url{http://doug-pc.itp.ucsb.edu/online/gravity_c99/schoen/}.

\bibitem{CutlerWald}
C.~Cutler and R.M. Wald, \emph{{Existence of radiating
Einstein-Maxwell
  solutions which are $C\sp{\infty}$ on all of ${\cal I}\sp+$ and ${\cal
  I}\sp-$}}, Class.\ Quantum Grav. \textbf{6} (1989), 453--466.

\bibitem{Dain:2001kn}
S.~Dain, \emph{Initial data for stationary space-times near
space-like
  infinity}, Class.\ Quantum Grav. \textbf{18} (2001), 4329--4338,
  gr-qc/0107018.

\bibitem{Damour:schmidt}
T.~Damour and B.~Schmidt, \emph{Reliability of perturbation theory
in general
  relativity}, Jour.\ Math.\ Phys. \textbf{31} (1990), 2441--2453.

\bibitem{Friedrich}
H.~Friedrich, \emph{On the existence of n--geodesically complete
or future
  complete solutions of {E}instein's field equations with smooth asymptotic
  structure}, Commun. Math. Phys. \textbf{107} (1986), 587--609.

\bibitem{HelmutJDG}
\bysame, \emph{On the global existence and the asymptotic behavior
of solutions
  to the {E}instein --- {M}axwell --- {Y}ang--{M}ills equations}, Jour.\ Diff.\
  Geom. \textbf{34} (1991), 275--345.

\bibitem{Friedrich:2002ru}
\bysame, \emph{Spin-2 fields on {Mi}nkowski space near space-like
and null
  infinity},  (2002), gr-qc/0209034.

\bibitem{GoldbergSoteriou}
J.N. Goldberg and C.~Soteriou, \emph{{Canonical general relativity
on a null
  surface with coordinate and gauge fixing}}, Class.\ Quantum Grav. \textbf{12}
  (1995), 2779--2797.

\bibitem{KlainermanNicolo}
S.~Klainerman and F.~N{icol\`o}, \emph{On local and global aspects
of the
  {C}auchy problem in general relativity}, Class.\ Quantum Grav. \textbf{16}
  (1999), R73--R157.

\bibitem{penrose:scri}
R.~Penrose, \emph{Zero rest-mass fields including gravitation},
Proc.\ Roy.\
  Soc.\ London \textbf{A284} (1965), 159--203.

\bibitem{SchoenYaubh}
R.~Schoen and S.-T. Yau, \emph{{The existence of a black hole due
to
  condensation of matter}}, Commun.\ Math.\ Phys. \textbf{90} (1983), 575--579.

\bibitem{TodCargese}
P.~Tod, talk given at the Carg\`ese Summer School, August 2002,
online at
  \url{fanfreluche.math.univ-tours.fr}.

\bibitem{winicour:logarithmic}
J.~Winicour, \emph{Logarithmic asymptotic flatness}, Found. Phys.
\textbf{15}
  (1985), 605--615.

\end{thebibliography}
\end{document}